# Fast and accurate Fe–H machine-learning interatomic potential for elucidating hydrogen embrittlement mechanisms

Kazuma Ito *

*Advanced Technology Research Laboratories, Nippon Steel Corporation, 20-1, Shintomi, Futtsu city, Chiba 293-8511, Japan*

* Corresponding author.

E-mail address: ito.nn3.kazuma@jp.nipponsteel.com (K. Ito)

**Abstract**

Understanding the mechanisms of hydrogen embrittlement (HE) is essential for advancing next-generation high-strength steels, thereby motivating the development of highly accurate machine-learning interatomic potentials (MLIPs) for the Fe–H binary system. However, the substantial computational expense associated with existing MLIPs has limited their applicability in practical, large-scale simulations. In this study, we construct a new MLIP within the Performant Implementation of the Atomic Cluster Expansion (PACE) framework, trained on a comprehensive HE-related dataset generated through a concurrent-learning strategy. The resulting potential achieves density functional theory–level accuracy in reproducing a wide range of lattice defects in α-Fe and their interactions with hydrogen, including both screw and edge dislocations. More importantly, it accurately captures the deformation and fracture behavior of nanopolycrystals containing hydrogen-segregated general grain boundaries—phenomena not explicitly represented in the training data. Despite its high fidelity, the developed potential requires computational resources only several tens of times greater than empirical potentials and is more than an order of magnitude faster than previously reported MLIPs. By delivering both a high-precision and computationally efficient potential, as well as a generalizable methodology for constructing such models, this study significantly advances the atomic-scale understanding of HE across a broad range of metallic materials.

# 1. Introduction

Green hydrogen, produced sustainably from water using renewable energy, is widely regarded as a key energy carrier for achieving a decarbonized society. However, ensuring the reliability of structural materials throughout its production, transport, storage, and utilization remains a major technological bottleneck [1]. Among the associated challenges, hydrogen embrittlement (HE) is particularly critical, as it severely degrades the ductility and toughness of metals, potentially leading to premature failure or catastrophic fracture [2-4]. HE has been documented in a broad range of structural materials, including ferritic steels [5–8], martensitic steels [5-9], austenitic stainless steels [10-12], nickel alloys [13-15], aluminum alloys [16-18], and titanium alloys [19, 20].

Although increasing the strength of metals is desirable for improving hydrogen storage efficiency and enabling structural optimization, it typically exacerbates susceptibility to HE. For instance, hydrogen-induced failure in steels becomes pronounced when the tensile strength exceeds approximately 1.2 GPa [21]. In martensitic steels, a representative category of high-strength steels, HE often manifests as quasi-cleavage and intergranular fracture, with severe grain boundary (GB) cracking observed at elevated hydrogen concentrations [22]. Notably, the GBs that fail preferentially are typically "general" GBs lacking crystallographic symmetry, prompting extensive investigations into HE mechanisms associated with such boundaries [7, 23-27].

The hydrogen-enhanced decohesion (HEDE) mechanism was initially proposed to explain GB-related HE [28]; however, it does not fully account for several key experimental findings, including quasi-cleavage features and dislocation pile-up detected beneath intergranular fracture surfaces. To address these limitations, the hydrogen-enhanced-plasticity-mediated decohesion mechanism was introduced [29], integrating HEDE with hydrogen-enhanced local plasticity (HELP) [30]. This combined framework suggests that both GB weakening and dislocation accumulation in the vicinity of GBs contribute to fracture. Recent studies have further highlighted that the ease of crystallographic plastic relaxation at a GB is a decisive factor governing its susceptibility to HE [24]. Nevertheless, directly observing hydrogen distribution near GBs remains experimentally challenging [31], leaving the atomic-scale mechanisms underlying GB-related HE only partially understood.

Computational materials science has recently emerged as a powerful tool for investigating HE [32-46]. However, density functional theory (DFT) calculations are constrained by their high computational cost, making it challenging to analyze general GBs and their associated deformation and fracture behaviors. Consequently, several studies have employed empirical interatomic potentials to examine HE accompanied by GB cracking [47]. Despite their computational efficiency, these empirical potentials raise serious accuracy concerns. For example, embedded-atom method (EAM) potentials widely used for Fe–H systems [48-51] significantly underestimate the energies of symmetric tilt GBs [52], fail to correctly capture the relative energies associated with core structural changes in screw dislocations [52], and exhibit unphysical behavior near crack tips [53-55].

In parallel, high-accuracy machine-learning interatomic potentials (MLIPs) have been developed for a variety of material systems [56-58]. For the Fe–H binary system, MLIPs based on the Behler–Parrinello neural network potential (BNNP) [59, 60] and the Deep Potential (DP) framework [61, 62] have been proposed, addressing many of the limitations inherent to conventional EAM potentials [60, 62]. Most recently, the Fe–H binary datasets used to construct these MLIPs were extended to develop an Fe–C–H ternary MLIP [63]. Nonetheless, existing Fe–H binary MLIPs exhibit significantly reduced accuracy when applied to atomic environments not included in their training sets—such as symmetric tilt GBs [64]—and fail to reliably reproduce hydrogen segregation behavior at general GBs [65].

To overcome these limitations, the authors previously developed a high-fidelity Fe–H MLIP within the BNNP framework capable of analyzing HE behavior at GBs, including general GB configurations [65]. This model was trained on a comprehensive dataset generated through a concurrent-learning approach [66], covering a wide spectrum of atomic environments associated with HE. Despite its excellent accuracy, its application in large-scale HE simulations has been limited due to its high computational cost—several hundred times that of EAM [65]. Moreover, the model has not been sufficiently validated for the complex atomic configurations that arise in realistic HE scenarios, raising concerns regarding its reliability in extrapolative regions. In such cases, MLIPs may not only lose their predictive accuracy but can also exhibit severe degradation [67], potentially leading to nonphysical behavior.

In this study, we address these challenges by employing the comprehensive Fe–H training dataset [65] to develop a new MLIP within the Performant Implementation of the Atomic Cluster Expansion (PACE) framework [68, 69], which provides an excellent balance between accuracy and computational efficiency [70]. The resulting potential accurately reproduces fundamental properties of α-Fe, formation energies of lattice defects, and hydrogen–defect interactions with DFT accuracy.

Notably, it also captures hydrogen interactions with general GBs, screw dislocations, and recently reported DFT-calculated edge dislocations [71], despite these configurations not being explicitly included in the training data. We further validate the model's accuracy for complex, large-scale systems—containing more than five million atoms—using extrapolation-grade–based evaluation [52, 72], demonstrating its ability to reproduce the deformation and fracture behaviors of hydrogen-free and hydrogen-segregated nanopolycrystals. Despite its superior accuracy, the computational cost of this MLIP remains only several tens of times that of EAM and more than an order of magnitude lower than that of conventional MLIPs, establishing it as a highly practical and powerful tool for advancing the mechanistic understanding of HE.

The remainder of this paper is organized as follows. Section 2 describes the construction of the MLIP and the methodologies employed for accuracy validation. Section 3 presents the performance of the MLIP for HE-relevant atomic environments and discusses the implications of these findings. Finally, Section 4 presents the conclusions.

## 2. Methods

This section describes the development of the Fe–H binary MLIP (Section 2.1), the procedures employed to evaluate its accuracy against DFT and assess its computational efficiency (Section 2.2), and the extrapolation-grade–based validation performed for complex atomic configurations that arise during the uniaxial deformation and fracture of hydrogen-free and hydrogen-segregated α-Fe nanopolycrystals (Section 2.3).

### 2.1. Construction of the Fe–H Binary MLIP

The training dataset comprised 36,438 atomic structures (1,549,196 atomic environments) generated using a concurrent-learning strategy designed to efficiently and comprehensively sample configurations relevant to HE. These included generalized stacking faults, surfaces, symmetric tilt GBs, vacancy clusters, and their hydrogen-containing counterparts (Table 1). Further details of the dataset construction and sampling procedure are provided in our previous work [65].

**Table 1. Training datasets for the Fe–H machine learning interatomic potential.** The training data include an initial dataset composed of six subsets and a DP-GEN dataset obtained from concurrent learning, containing eight subsets.

| Subsets | Number of structures |
|---|---|
| a. Initial dataset | |
| 1. Equilibrated Fe with dilute H | 144 |
| 2. Perturbed Fe with dilute H | 595 |
| 3. Strained Fe with 1 H | 99 |
| 4. H-vacancy clusters | 124 |
| 5. 2H in neighboring TISs | 207 |
| 6. Single H atom and $H_2$ molecules | 24 |
| All initial datasets | 270 |
| b. DP-GEN dataset | |
| 1. Fe with dilute H | 1212 |
| 2. Fe with high concentration H | 2088 |
| 3. H atoms in vacancy | 901 |
| 4. H atoms on surface | 1408 |
| 5. Generalized stacking faults with H atoms | 1081 |
| 6. Perturbed tilt grain boundaries with H atoms | 4924 |
| 7. Self-interstitial atom ($\langle 111 \rangle$, $\langle 110 \rangle$, $\langle 100 \rangle$ dumbbell) | 5591 |
| 8. H atoms in vacancy cluster | 16531 |
| All datasets | 36438 |

In the previous study [65], three MLIP frameworks—BNNP [59], the Moment Tensor Potential (MTP) [73], and DP [74]—were evaluated, with BNNP ultimately selected for offering the best balance between accuracy and computational cost on Fugaku, a CPU-based supercomputing system. More recent benchmarks using Al–Cu–Zr and Si–O datasets have demonstrated that the PACE framework can outperform both BNNP and MTP in terms of the trade-off between fitting accuracy and computational efficiency [70], with similarly strong performance reported for α-Fe [75]. Moreover, PACE has been successfully applied in the development of Fe–O MLIPs [76]. Motivated by these findings, the present study employs the PACE framework to construct a Fe–H MLIP that offers an improved balance between accuracy and computational speed. However, it should be noted that prior PACE-based Fe MLIPs [75, 76] did not evaluate model reliability for the complex systems considered in this study—such as general GBs, their deformation and fracture behavior, or the effects of solute elements on these processes.

In the atomic cluster expansion (ACE) formalism, the local energy is expressed as a function of atomic properties, as follows [77]:

$$E_i = \mathcal{F}\left(\varphi_i^{(1)}, \dots, \varphi_i^{(P)}\right) \tag{1}$$

where $\varphi_i^{(P)}$ can be expanded as follows:

$$\varphi_i^{(p)} = \Sigma_{v=1}^{v_{\max}} c_v^{(p)} B_{iv} \tag{2}$$

$c_v^{(p)}$ denote the expansion coefficients to be fitted by the regression algorithm. $B_{iv}$ denotes the basis constructed by the atomic cluster expansion, where the permutation, reflection, and rotation invariants are incorporated [77]. The **B** basis is efficiently reconstructed through the multiplication of generalized Clebsch–Gordan coefficients with the permutation-invariant **A** basis functions. Among the various formulations proposed for $\mathcal{F}$, this study adopts a nonlinear expansion of the local density [70]:

$$E_i = \varphi + \sqrt{\varphi} + \varphi^2 + \varphi^{0.75} + \varphi^{0.25} + \varphi^{0.875} + \varphi^{0.625} + \varphi^{0.375} + \varphi^{0.125} \tag{3}$$

For the expansion of atomic properties, we employed 544 basis functions with a total of 4982 fitting parameters. Bessel functions were used to construct the radial basis, and cutoff radii were set to 6.0 Å for Fe–Fe interactions and 4.5 Å for all other interactions. These choices enabled accurate fitting of diverse lattice defects, H–defect interactions, and H–H interactions. Training was conducted using the Pacemaker code [68]. The weighting of forces and energies was set to Auto within Pacemaker, yielding an optimized value of 0.52. Parameter optimization was performed using the BFGS algorithm. The resulting root mean square errors (RMSEs) for energies and forces were 5.85 meV/atom and 87.26 meV/Å, respectively—comparable to the accuracy previously achieved with the BNNP-based MLIP trained on the same dataset [65] (4.09 meV/atom, 109.16 meV/Å)

## 2.2 Accuracy verification against DFT and evaluation of computational efficiency

To assess the accuracy of the developed MLIP, we computed fundamental properties of α-Fe and formation energies of representative lattice defects, and compared them with corresponding DFT results. Hydrogen–defect interactions were also evaluated, including H solution energies under isotropic and uniaxial strains, segregation energies at low-index surfaces, symmetric tilt GBs, screw dislocations, edge dislocations, and general GBs. Detailed computational procedures are described in Refs. [65, 71].

Computational efficiency was evaluated following the methodology of Ref. [52]. α-Fe supercells of varying sizes were equilibrated at 300 K under the NVT ensemble, and the wall-clock time per atom per molecular dynamics (MD) step was recorded. Because the number of hydrogen atoms is typically much smaller than that of Fe atoms in HE-related simulations, the computational cost is effectively dominated by the Fe atoms; thus, Fe-only supercells were used for performance benchmarking. Consistently, throughout the simulations conducted in this study, the computational efficiency was primarily determined by the number of Fe atoms.

All simulations were performed on the supercomputer Fugaku, whose compute nodes contain 48-core A64FX processors interconnected through the TofuD network. Two nodes were utilized to ensure sufficient memory capacity.

In addition to the MLIP developed in this study, we also evaluated two BNNP-based models: the BNNP-Ito potential [65], which was trained on the same dataset used here, and the independently developed BNNP-Meng potential [60]. A DP model trained on the identical dataset in Ref. [60] has also been constructed for GPU-accelerated computation [62]; however, because the present benchmarks were performed on CPU-only architectures—and given that previous work demonstrated that BNNP surpasses DP both in fitting accuracy and in predicting the defect properties and H–defect interactions examined in this study [62]—the BNNP models were selected for comparison istead.

Furthermore, we evaluated the empirical EAM potential developed by Song et al. [48, 51], which is widely employed in HE-related MD simulations. Although several other empirical Fe–H potentials have been proposed recently [78, 79], a recent benchmark study [80] identified the Song potential as providing the highest overall accuracy among them. The DFT reference data used for validation were identical to those employed in our previous work [65].

## 2.3. Accuracy verification of the Fe–H MLIP for hydrogen-free and hydrogen-segregated nanopolycrystals under deformation and fracture

Using the developed MLIP, uniaxial tensile simulations were performed on both hydrogen-free and hydrogen-segregated nanopolycrystals, followed by a detailed analysis of the resulting atomic configurations. The accuracy of the MLIP was then evaluated for these structures formed during deformation. Although the effects of hydrogen on the deformation and fracture behavior of nanopolycrystals are described, the primary aim here is to evaluate the MLIP's predictive performance for atomic configurations relevant to GB-mediated HE. Despite its relatively high efficiency, the MLIP—like any MD-based approach—remains constrained by the computational cost of resolving individual atomic motions. Consequently, the grain sizes modeled in this study are limited to the nanometer scale, far smaller than those in typical steels. Furthermore, the strain rates achievable in MD simulations are approximately ten orders of magnitude higher than those at which macroscopic HE phenomena occur [81]. Therefore, macroscopic indicators such as stress–strain curves from these simulations are not expected to directly match experimental trends. Nevertheless, the local atomic configurations generated under these conditions are expected to be representative of those associated with GB-related HE.

The initial nanopolycrystal structures were identical to those used previously in uniaxial tensile simulations of pure α-Fe employing an MTP [64] (Fig. 1(a)). This MTP has been shown to reliably capture the deformation and fracture behavior of pure α-Fe nanopolycrystals [64], enabling the MLIP predictions to be benchmarked against established results. The initial configurations consisted of cubic nanopolycrystals with 16 grains of random crystallographic orientations, generated by Voronoi tessellation using Atomsk [82]. To account for variability arising from grain seed (relative positions and orientations), five distinct seeds were prepared. To further examine the influence of grain size, the edge lengths of the initial structures were set to 10, 15, 20, 25, 30, 35, and 40 nm for each seed, yielding a total of 35 configurations with corresponding average grain sizes of 4.8, 7.3, 9.7, 12.1,

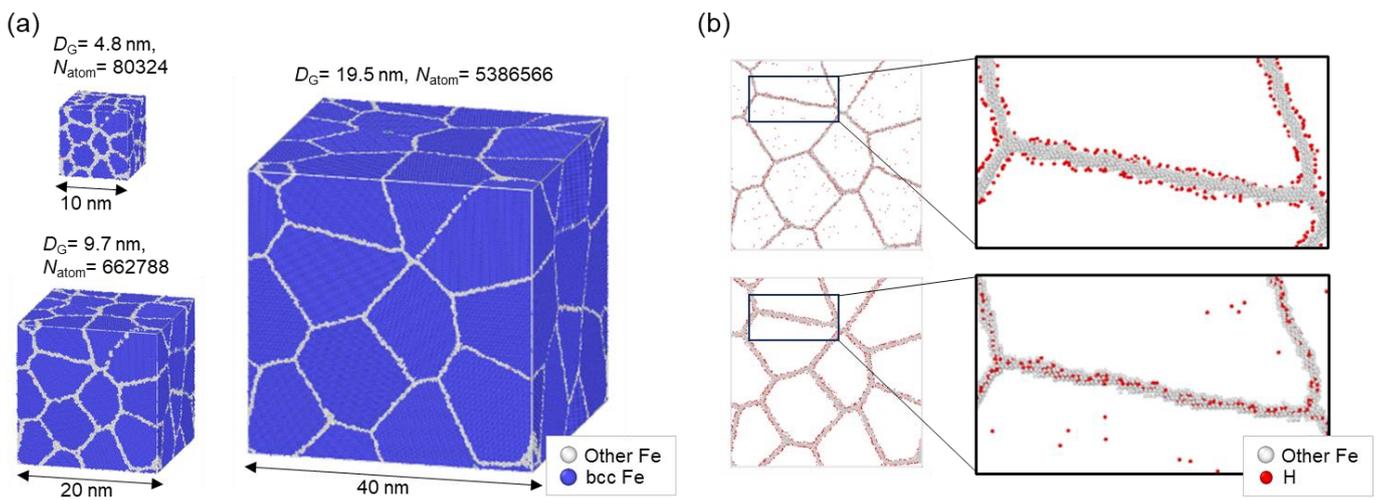

**Fig. 1. Atomic structures of the nanocrystalline models.** (a) Example of a relaxed atomic configuration of pure Fe nanocrystals. The average grain size $D_G$ and the number of atoms $N_{atom}$ for each model are indicated. (b) Evolution of the hydrogen distribution during annealing used to generate the hydrogen-segregated nanocrystalline models.

14.6, 17.0, and 19.5 nm.

For consistency, the relaxation conditions for the nanopolycrystals were identical to those used in the α-Fe nanopolycrystal tensile simulations with the MTP, and were also aligned with the EAM-based procedure of Jeon et al. [83]. The initial structures were first optimized using the conjugate gradient method, followed by annealing at 300 K for 200 ps in a Nosé–Hoover isothermal–isobaric (NPT) ensemble. Periodic boundary conditions were applied in all directions to eliminate surface effects.

GBs with hydrogen segregation were prepared by annealing nanopolycrystals in which hydrogen atoms had been inserted into the previously relaxed GB structures, followed by an additional 200 ps of relaxation at 300 K to achieve the target GB hydrogen concentrations. Since three hydrogen concentration levels were considered, it was impractical to perform this procedure for all nanopolycrystals. Therefore, from the five α-Fe nanopolycrystal seeds, the seed whose flow stress most closely matched the overall average was selected. Two grain sizes, 4.8 and 19.5 nm, were chosen for analysis, as these sizes—as shown later—exhibit distinct dominant plastic deformation mechanisms.

In this analysis, we sought to isolate the effect of segregated hydrogen as clearly as possible. To minimize changes to the underlying GB structure, hydrogen atoms were first inserted, after which the system was annealed at 300 K (Fig. 1(b)). To promote segregation during annealing, hydrogen atoms were initially placed at tetrahedral interstitial sites located 2.5–5.0 nm from the GB center. The number of hydrogen atoms was selected such that, assuming complete segregation to the boundary, the hydrogen concentration—defined as the number of H atoms divided by the number of Fe atoms within 2.5 nm of the GB—reached 10.0, 20.0, or 30.0 at.%. The resulting models are referred to as 10 at.% H, 20 at.% H, and 30 at.% H, respectively. Following hydrogen insertion, each system was held for 1 ns to allow diffusion-driven segregation and the formation of physically realistic boundary hydrogen distributions. For consistency in evaluating hydrogen effects, hydrogen-free α-Fe nanopolycrystals used for comparison were subjected to the same additional 1 ns annealing.

A concentration of 10 at.% approximates the equilibrium GB hydrogen coverage of 2.3–2.9 atom/nm² measured by Sato et al. [84] in electrolytically charged pure Fe. However, GBs vulnerable to HE are expected to accumulate hydrogen beyond these equilibrium levels, driven by factors such as deformation-induced local strain [64] and dislocation-mediated hydrogen transport [29]. Therefore, we additionally considered the 20.0 and 30.0 at.% cases. Notably, 20.0 at.% corresponds to the hydrogen segregation level at which intergranular fracture occurred without dislocation emission in a previous BNNP-based bicrystal study involving a general GB [65].

The resulting nanopolycrystals were subjected to uniaxial tensile deformation at a strain rate of $5 \times 10^8$ s$^{-1}$ while maintaining zero lateral stress. These conditions match those used in prior MTP-based tensile simulations of $\alpha$-Fe nanopolycrystals, enabling direct comparison. This strain rate represents a widely adopted compromise in MD simulations of nanopolycrystals, mitigating excessive stress overshoot associated with extremely high strain rates while avoiding the prohibitive cost of slower deformation [85-88]. This strain rate is also comparable to those typically employed in MD investigations of hydrogen-induced GB fracture [35, 40, 89]. The timestep was set to 1.0 fs, a standard choice in HE-related tensile simulations [85-88].

Directly benchmarking the MLIP against DFT for the atomic structures generated during nanopolycrystal deformation is infeasible due to the extreme computational cost. Instead, we adopted the validation strategy previously used to assess MTP accuracy in complex α-Fe and W systems [52, 90]. Specifically, we evaluated the extrapolation grade [72]—a metric derived from active-learning theory that quantifies the similarity of local atomic environments to those present in the training dataset—to determine whether each atomic environment lies within the interpolation regime, where the potential is expected to provide reliable predictions. Although originally developed for MTPs [72], this metric is equally applicable to PACE-type potentials [68].

In this framework, atomic environments with extrapolation grades between 0 and 1 are classified as belonging to the interpolation region, where the MLIP is expected to achieve high accuracy. Grades between 1 and 2 correspond to the accurate extrapolation regime, while grades from 2 to 10 indicate reliable extrapolation. In active-learning workflows, structures containing environments in this latter range are typically selected for DFT labeling, as they represent insufficiently sampled regions of configuration space. Accordingly, if the extrapolation grades remain below 2, the MLIP can be regarded as highly accurate for the system of interest, without requiring additional training data. Notably, extrapolation grade has been shown to correlate strongly with the deviation of MLIP predictions from DFT [52]. Additional details can be found in Ref. [72]. For both pure α-Fe nanopolycrystals and those containing hydrogen-segregated GBs under uniaxial tension, atomic configurations were sampled every 1 ps during deformation, and extrapolation grades were computed for all atoms in each configuration.

## 2.4. Details of the Molecular Dynamics Simulations

LAMMPS [91] was used for the interatomic-potential-based simulations. OVITO [92] was used for visualization and analysis of the atomic structures. The local crystal structure of each atom was identified using the Polyhedral Template Matching (PTM) method [93]. The Dislocation Extraction Algorithm (DXA) [94] was used to visualize dislocations and quantify dislocation densities. Intragranular dislocation density was evaluated by applying DXA to atoms identified as bcc by PTM. Deformation twins were detected using PTM combined with grain segmentation, and the number of atoms within twins was used to compute the twin-atom fraction and the number of twinned regions. A surface mesh approach [95] was further applied to evaluate crack volume and crack counts.

## 3. Results and Discussion

### 3.1. Accuracy of the constructed Fe–H binary MLIP

As detailed in the Supplementary Materials, the PACE-based MLIP developed in this study reproduces the fundamental properties of α-Fe and the formation energies of representative lattice defects with accuracy comparable to previously reported MLIPs. The results also in good agreement with earlier DFT calculations [60] and with experimental data [96-98]. Moreover, the PACE accurately captures generalized stacking fault energies and the energetics of screw-dislocation cores [99, 100]—both critical to the plastic deformation behavior of α-Fe—with fidelity similar to that reported in prior studies [60, 62, 65].

Figure 2 shows the accuracy of the MLIP in describing interactions between hydrogen and various lattice defects in α-Fe. The PACE successfully reproduces hydrogen solution energies at interstitial sites under uniaxial and hydrostatic strain (Fig. 2(a)–(c)), which is essential for evaluating hydrogen stability during deformation and for characterizing its interactions with diverse lattice defects. Notably, the uniaxially strained octahedral site corresponds to the dominant segregation site at general GBs [33]. The effect of hydrogen on the generalized stacking fault energy on the (112) plane—relevant to twinning—was likewise well captured (Fig. 2(d)). The interactions between a monovacancy and hydrogen (Fig. 2(e)) and the segregation energies on low-index surfaces (Fig. 2(f)) also showed strong consistency with DFT. The PACE additionally reproduces the dependence of $H_2$ molecular energy on bond length (Fig. 2(g)). For hydrogen–hydrogen interaction energies in bulk α-Fe, the PACE exhibits better agreement with DFT than BNNP-Meng, accurately capturing the repulsive interaction between hydrogen atoms at neighboring sites (Fig. 2(h)). All of these results correspond to configurations explicitly included in the training dataset.

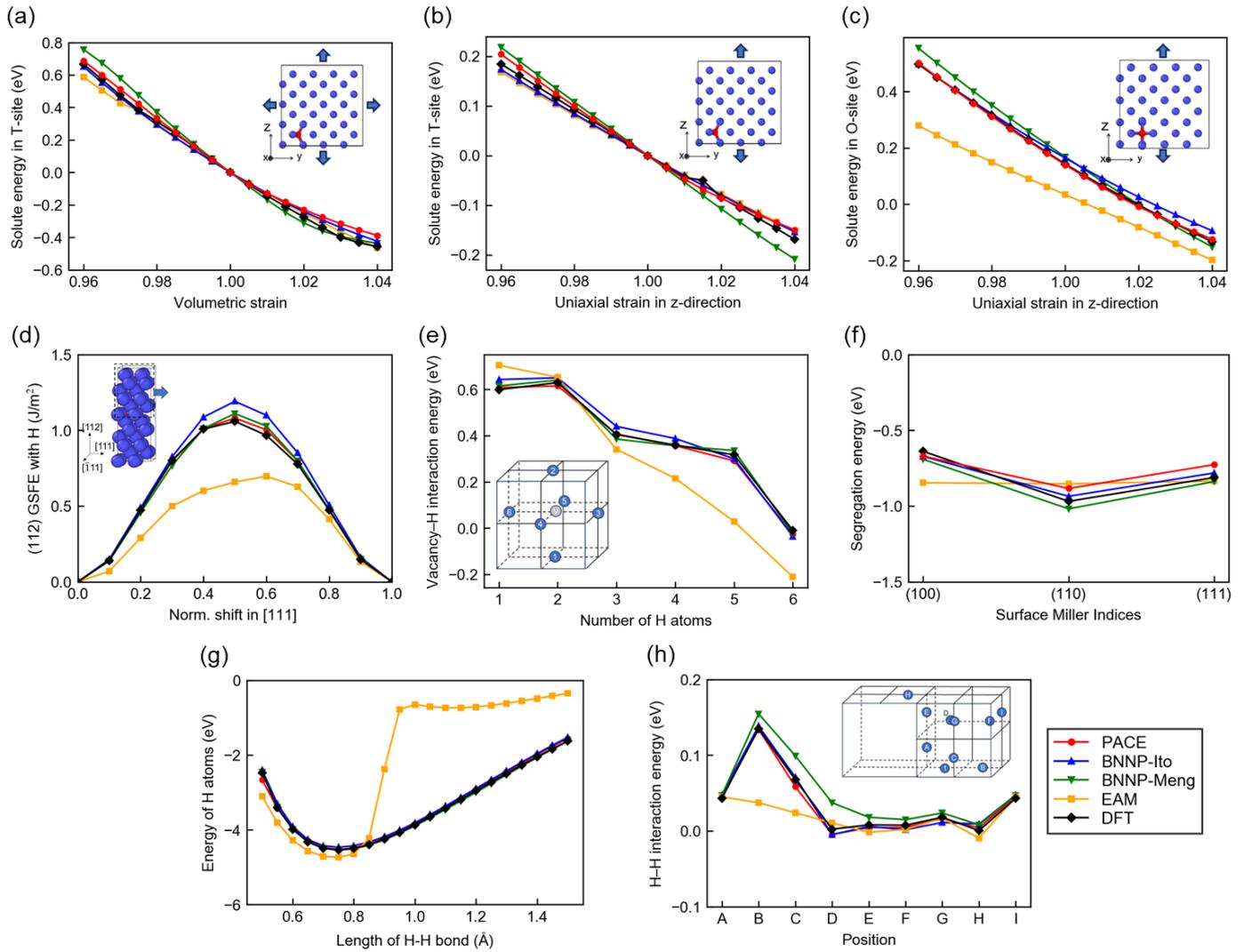

**Fig. 2. Accuracy of the constructed PACE-based MLIP for the Fe–H binary system.** (a) Hydrogen solution energy at tetrahedral sites in α-Fe under isotropic strain, and (b) under uniaxial strain. (c) Hydrogen solution energy at octahedral sites in α-Fe under uniaxial strain. (d) Generalized stacking fault (GSF) energy when hydrogen is placed at a tetrahedral site on the (112) GSF plane. (e) Interaction energy between a single vacancy and multiple hydrogen atoms. (f) Segregation energy of hydrogen at the most stable sites on low-index surfaces. (g) Bond-length dependence of the energy of a hydrogen molecule. (h) Interaction energy between hydrogen atoms dissolved at tetrahedral sites in bulk α-Fe. For comparison, results obtained from DFT, the BNNP developed by Ito et al. and Meng et al., and EAM are also shown. In (e) and (h), the hydrogen configurations corresponding to the horizontal axis are indicated.

The accuracy of the PACE in describing interactions between dislocations and hydrogen—an essential factor in understanding hydrogen-induced modifications to plasticity—is presented in Fig. 3. The PACE reproduces DFT interaction energies for screw dislocations with notably higher accuracy than BNNP-Meng (Fig. 3(a)), despite the latter explicitly incorporating screw-dislocation configurations into its training set, whereas our dataset does not. For edge dislocations, the accuracy of the PACE is slightly lower than that of BNNP-Ito but remains superior to BNNP-Meng (Fig. 3(b)).

Figure 4 summarizes the hydrogen segregation energies for symmetric tilt GBs and general GBs. The general GB structures were extracted from a nanocrystalline α-Fe model generated via Voronoi tessellation and relaxed using an MTP known to reproduce general GB structures with near-DFT accuracy [68]. As shown in Fig. 4(d), supercells cut from a 28.3-nm model were used for DFT calculations, with hydrogen atoms inserted near the center to minimize interactions through periodic

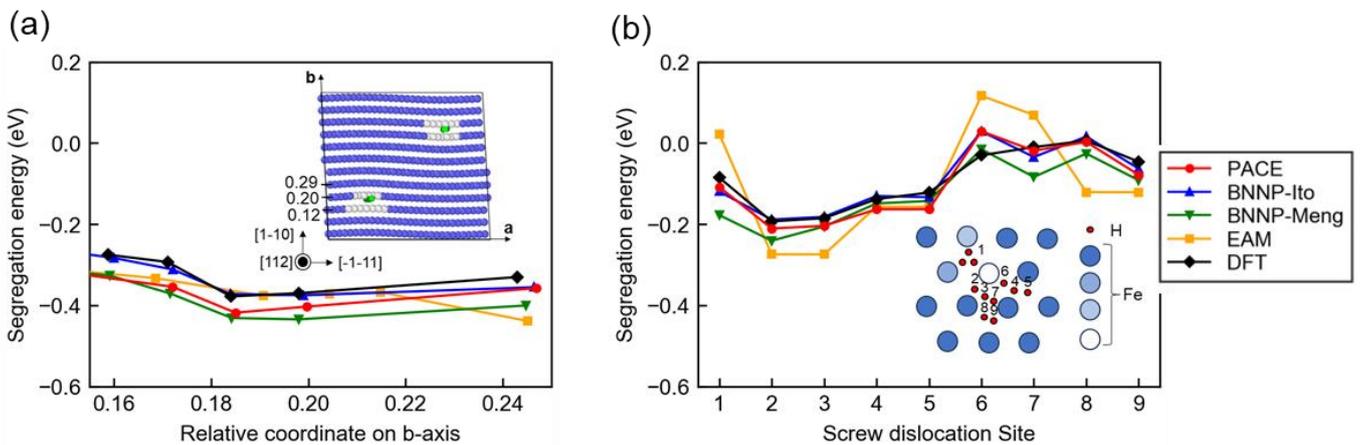

**Fig. 3. Accuracy of the constructed PACE-based MLIP for hydrogen segregation at dislocations.** (a) Hydrogen segregation energy at the $a_0/2\langle 111\rangle\{110\}$ edge dislocation, and (b) at the easy-core structure of the $a_0/2\langle 111\rangle\{110\}$ screw dislocation. For comparison, results obtained from DFT, the BNNP developed by Ito et al. and Meng et al., and EAM are also shown. In (a) and (b), the atomic structures of the respective dislocations and the hydrogen positions corresponding to the horizontal axis are indicated.

boundaries. For the MLIP-based calculations, candidate segregation sites were identified following the procedure in Ref. [33], and relaxed DFT geometries were used as initial configurations; atomic relaxation was then performed for both DFT and MLIP. Further computational details are available in Ref. [65]. Figures 4(a) and 4(b) show that the RMSE values for hydrogen segregation energies at eight symmetric tilt GBs and at general GBs are 0.046 eV and 0.039 eV, respectively—comparable to BNNP-Ito and better than BNNP-Meng. While segregation at symmetric tilt GBs was included in the training dataset, segregation at general GBs was not; thus, the strong agreement highlights the excellent transferability of the PACE model. A small number of sites exhibit deviations between DFT and the MLIP, corresponding to cases in which hydrogen migrates significantly during MLIP-based

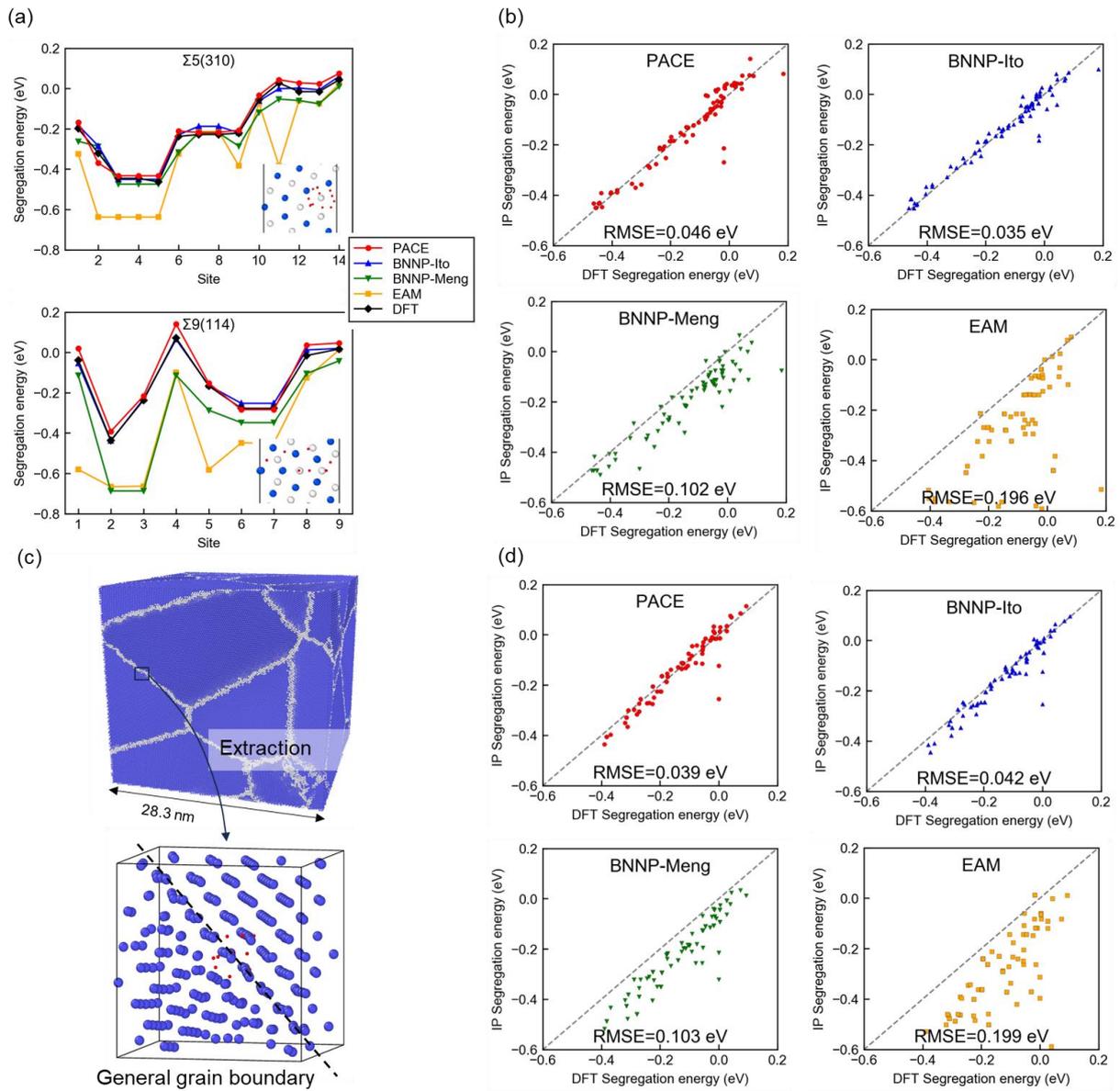

**Fig. 4. Accuracy of the constructed PACE-based machine learning interatomic potential for hydrogen segregation at grain boundaries.** (a) Hydrogen segregation energies at the Σ5(310) and Σ9(114) symmetric tilt grain boundaries. Atomic structures near the grain-boundary centers are shown, where blue and white spheres represent Fe atoms colored according to their coordinates along the out-of-plane (rotation-axis) direction, and red spheres denote hydrogen positions obtained from relaxed DFT calculations. Segregation sites at each grain boundary are numbered in order of increasing distance from the boundary center.
(b) Comparison of segregation energies at all sites in Σ3(112), Σ3(111), Σ5(210), Σ5(310), Σ9(114), Σ9(221), and Σ11(113) between DFT and the interatomic potential.
(c) Schematic of the method used to compute hydrogen segregation energies at general grain boundaries and an example of the computational cell. Blue spheres represent Fe atoms, and red spheres indicate the initial hydrogen positions used for segregation energy evaluation.
(d) Comparison of segregation energies for 85 sites across five extracted grain boundaries between DFT and the interatomic potential. In (b) and (d), results from the BNNP models of Ito et al. and Meng et al., and EAM, are also included for comparison, along with the root-mean-square error (RMSE) of the hydrogen segregation energies.

LAMMPS.

Figure 5 compares the computational cost of each interatomic potential. For sufficiently large systems, the per-atom computation time per timestep converges to a constant for all potentials (Fig. 5(a)). Figure 5(b) shows the computation time for a 250,000-atom α-Fe supercell (50 × 50 × 50), normalized to EAM (= 1). BNNP-Meng requires 837 × EAM, while BNNP-Ito is faster at 272 × EAM, consistent with typical MLIP costs. In contrast, PACE requires only 27 × EAM—over an order of magnitude faster than BNNP-Ito.

In summary, the PACE constructed from a comprehensive, efficiently sampled concurrent-learning dataset achieves accuracy comparable to the BNNP trained on the same data. It reproduces DFT values for bulk α-Fe properties, a wide range of lattice defects, and defect–hydrogen interactions. Moreover, it accurately describes hydrogen interactions with general GBs, screw dislocations, and edge dislocations—despite none of these configurations being explicitly included in the training data—demonstrating its strong transferability. At the same time, its computational cost is more than an order of magnitude lower than BNNP, underscoring its exceptional computational efficiency.

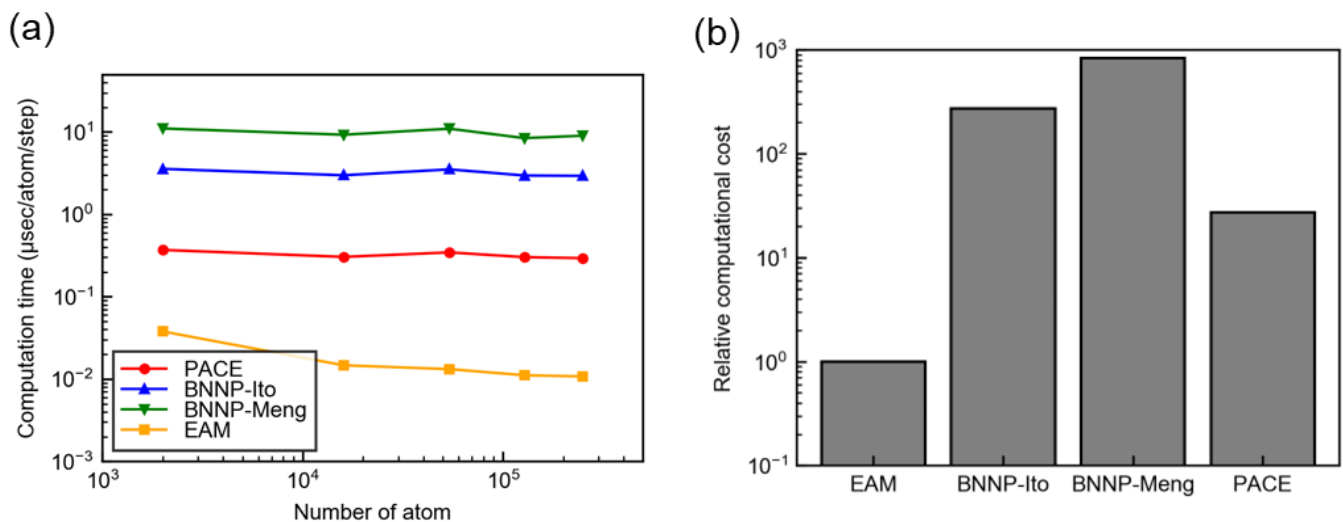

**Fig. 5. Computational speed of each interatomic potential.** (a) Computational cost per atom per timestep as a function of the number of atoms. (b) Relative computational cost per atom per timestep for each interatomic potential, normalized to that of EAM (=1), evaluated using a 50 × 50 × 50 α-Fe supercell containing 250,000 atoms.

## 3.2. Accuracy of the Fe–H MLIP for the deformation and fracture behavior of hydrogen-free and hydrogen-segregated nanopolycrystals

We first evaluated the ability of the constructed MLIP to reproduce the deformation and fracture behavior of pure α-Fe nanocrystals by comparing its predictions with high-fidelity reference data obtained using an MTP capable of accurately capturing general GB structures. Figure 6(a) presents representative stress–strain curves for nanocrystals with average grain sizes of 4.8, 9.7, and 19.5 nm. These curves correspond to random-seed configurations whose grain-size-dependent flow stresses most closely matched the ensemble average for N = 5. For all grain sizes, the tensile stress increased with strain until reaching a peak, followed by a gradual decay to a steady-state value. The peak stress marks the onset of plastic deformation and is known to increase under the high strain rates typically used in MD simulations, with the effect becoming more pronounced for larger grains. Consequently, when evaluating grain-size effects on nanocrystal deformation using MD, it is standard practice to quantify the mean flow stress over defined strain intervals, as the peak stress is highly sensitive to strain rate [101]. To avoid the influence of premature crack initiation at larger grain sizes, we computed the flow stress in the 3–7% strain range—where no significant crack nucleation occurred for any grain size—and in the 7–18% strain range following crack initiation. These strain intervals were selected in accordance with a previous MTP study [52]. The resulting flow stresses show excellent agreement with MTP values in both regimes (Fig. 6(b)), demonstrating the high fidelity of the developed MLIP in capturing the deformation and fracture behavior of pure α-Fe nanocrystals.

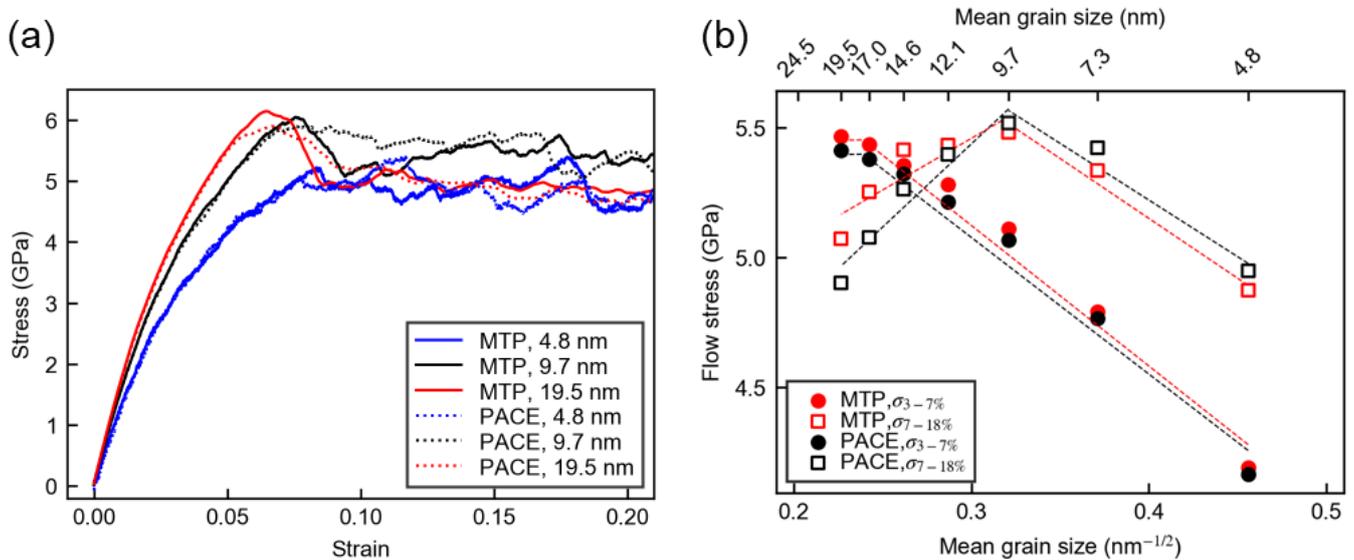

**Fig. 6. Dependence of the strength of pure α-Fe nanocrystals on the average grain size.**
(a) Stress–strain curves for nanocrystals with average grain sizes of 4.8, 9.7, and 19.5 nm. (b) Relationship between flow stress and grain size, averaged over five different nanocrystal seeds. Flow stress is shown for the strain range of 3–7%, where cracks were not prominently formed, and for the strain range of 7–18%, after crack initiation. For comparison, results calculated using the moment tensor-type MLIP from previous studies are also shown.

Prior MTP-based analyses [52] have clarified the origins of grain-size-dependent flow stress. In nanopolycrystals with average grain sizes of 19.7 nm (Hall–Petch regime), plastic deformation is governed primarily by dislocation and deformation twinning. In contrast, for grains averaging 4.8 nm (inverse Hall–Petch regime), GB-mediated mechanisms such as GB sliding and grain rotation dominate. Nanopolycrystals with intermediate grain sizes (e.g., 9.7 nm) display a coexistence of these deformation modes. In the following analysis, we therefore focus on nanocrystals with average grain sizes of 4.8 nm and 19.7 nm to investigate the effects of hydrogen segregation on deformation and fracture, identify the associated local atomic structures, and assess the predictive capability of the developed MLIP.

Figure 7 illustrates the evolution of atomic structures during uniaxial tension in a nanocrystal with an average grain size of 4.8 nm, together with the influence of hydrogen segregation. Figure 8 presents the corresponding stress–strain responses and quantitative assessments of atomic structural changes and hydrogen effects. At this grain size, GB sliding is the dominant deformation mechanism, and intragranular dislocations are essentially absent. In the hydrogen-free condition, grain shapes undergo substantial rearrangement as strain increases (Fig. 7). In contrast, in the presence of hydrogen, the degree of grain-shape evolution is markedly reduced (Fig. 7). Moreover, the maximum stress increases systematically with increasing hydrogen content (Fig. 8(a)). These results indicate that hydrogen segregation at GBs suppresses GB sliding.

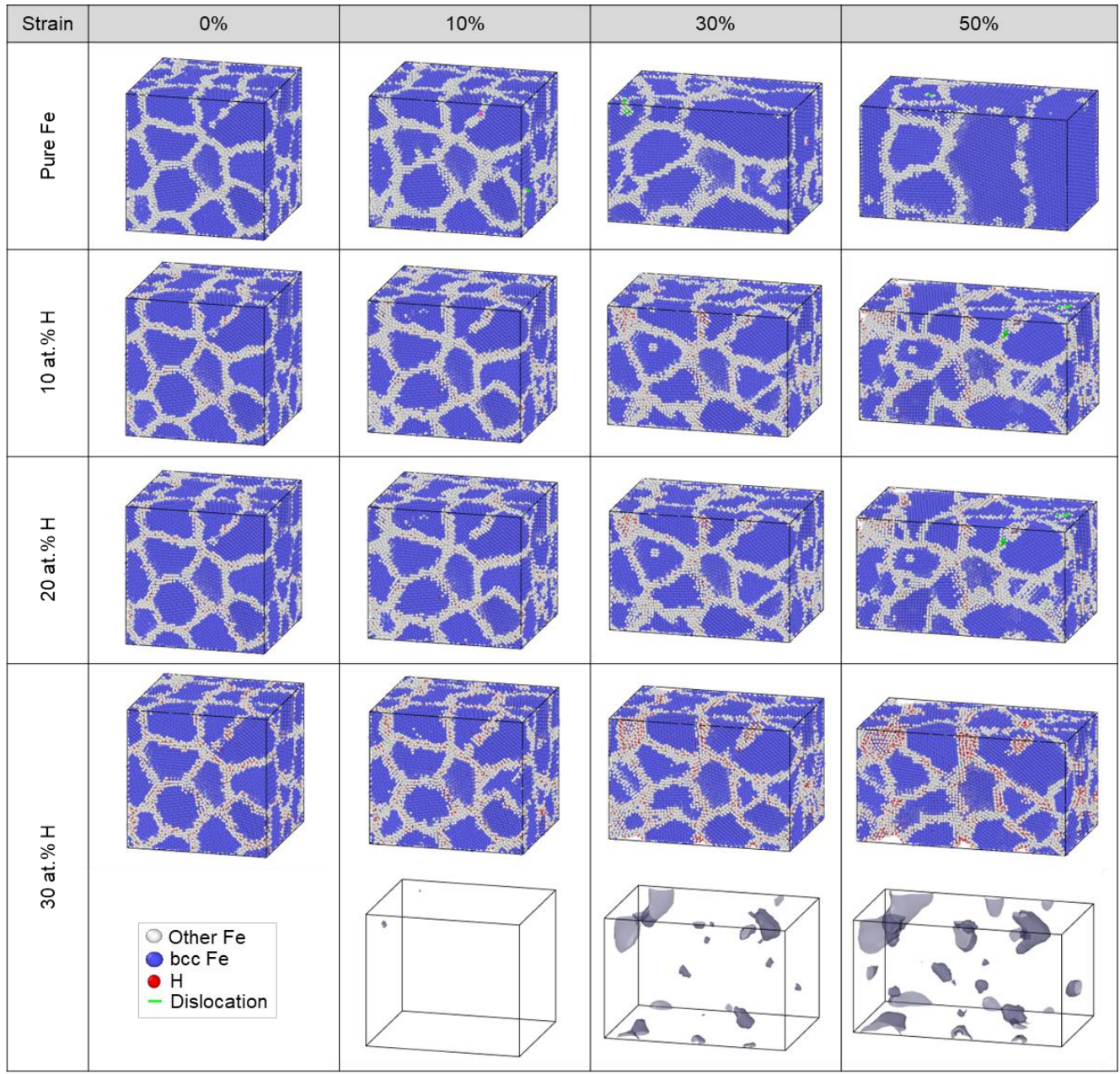

**Fig. 7. Atomic structural evolution in α-Fe nanocrystals with an average grain size of 4.8 nm under uniaxial tension and the effects of hydrogen.** Atomic crystal structures are identified using PTM, and dislocations extracted via DXA are also shown. The cases of 10 at.% H, 20 at.% H, and 30 at.% H correspond to the hydrogen concentrations at the grain boundaries. For the 30 at.% H case, an example of the crack morphology obtained using the surface-mesh method is additionally displayed.

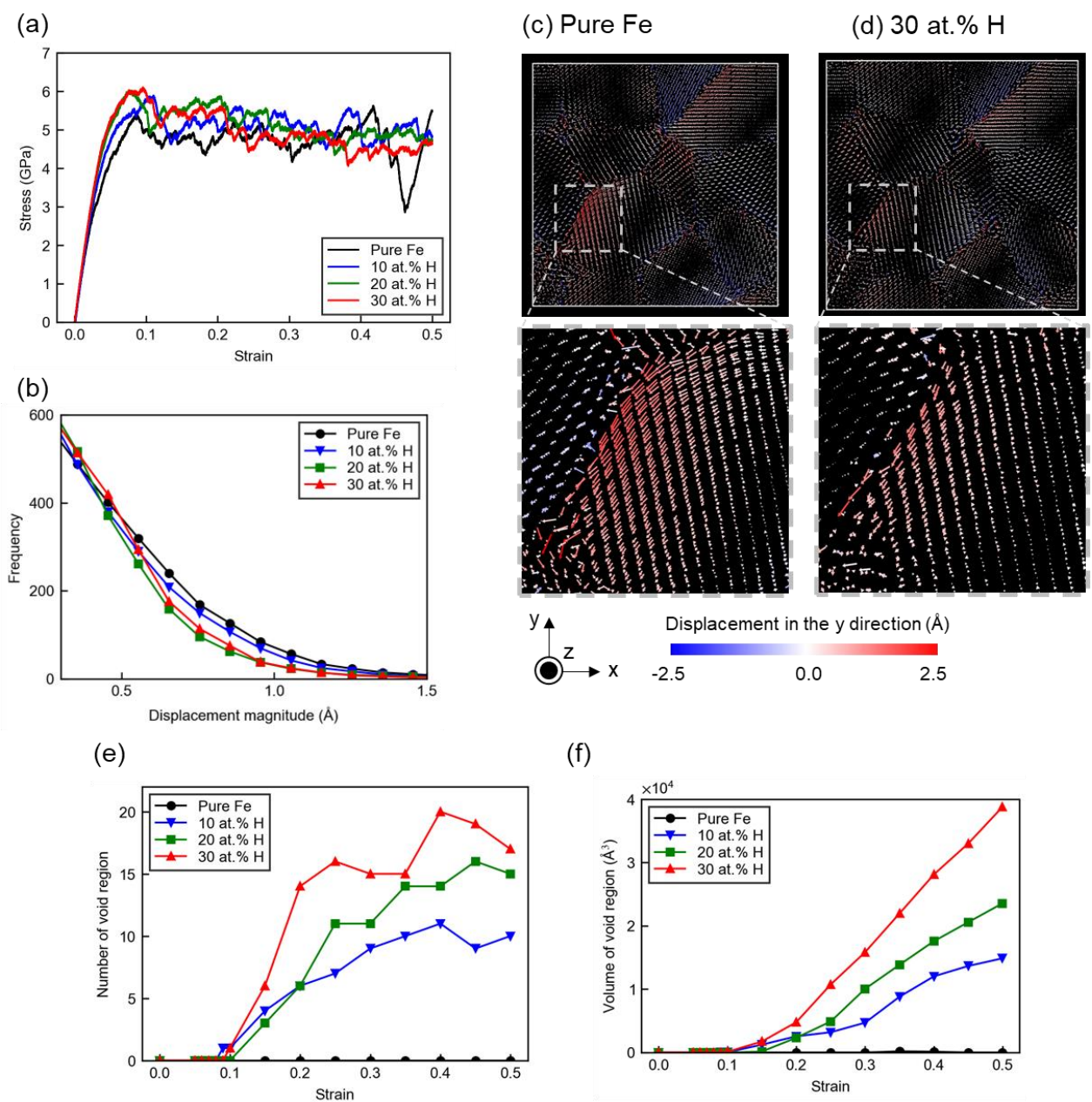

**Fig. 8. Quantitative analysis of atomic structural evolution in α-Fe nanocrystals with an average grain size of 4.7 nm under uniaxial tension and the effects of hydrogen.** (a) Stress–strain curves. (b) Histogram of atomic displacement magnitudes from 0% to 8% strain, corresponding to the vicinity of the maximum stress. (c) and (d) Atomic displacements from 0% to 8% strain for the hydrogen-free case and the 30 at.% H case, respectively. Arrow colors indicate displacements perpendicular to the tensile direction (the y-direction in the figure). Evolution with increasing strain of (e) the number of cracks and (f) crack volumes.

To quantify this effect, Fig. 8(b) presents a histogram of atomic displacements from 0% to approximately 8% strain, corresponding to the vicinity of the maximum stress. Figures 8(c) and 8(d) show the spatial distribution of atomic displacements over the same strain range for the hydrogen-free and 30 at.% H conditions, respectively. Arrow colors represent displacements perpendicular to the tensile direction (the y-direction). In all cases, atoms near GBs exhibit large displacements, confirming that deformation is governed by GB-mediated processes. Furthermore, the enlarged views in Figs. 8(c) and 8(d) clearly demonstrate that GB sliding is significantly reduced in the 30 at.% H case compared with the hydrogen-free case. Consistently, the histogram in Fig. 8(b) shows a decrease in the number of atoms undergoing large displacements with increasing hydrogen content. However, the similarity between the 20 at.% H and 30 at.% H cases suggests that the suppression of GB sliding saturates at a GB hydrogen concentration of roughly 20 at.%. These results establish that hydrogen segregation suppresses GB-mediated deformation and increases the flow stress in the low-strain regime.

Turning to the influence of hydrogen on crack formation, no cracks appear up to 50% strain in the hydrogen-free nanocrystal, whereas both the number and volume fraction of cracks increase with hydrogen content (Figs. 8(e),(f)). Notably, both quantities continue to rise between 20 at.% H and 30 at.% H. Since GB sliding suppression saturates at around 20 at.% H, these trends indicate that, beyond inhibiting GB-mediated plasticity, hydrogen segregation actively promotes crack initiation and propagation. Consequently, in the high-strain regime beyond the peak stress, the flow stress decreases with increasing hydrogen content (Fig. 8(a)).

Figure 9(a) illustrates the strain dependence of the mean and standard deviation of the extrapolation grade for all atoms (approximately 80,000 Fe atoms and up to 7,000 H atoms) during tensile deformation. Figures 9(b) and 9(c) show the strain dependence of the number of atoms with extrapolation grades in the ranges 1–2 and ≥2, respectively. The mean extrapolation grade remains nearly constant at approximately 0.4 across all strains, corresponding to the "interpolation" region where high accuracy is expected (Fig. 9(a)). In the absence of hydrogen, all atoms fall within this interpolation region (Figs. 9(b) and 9(c)). Furthermore, none of the hydrogen-free nanocrystalline samples—across all grain sizes and seeds—contained atoms with an extrapolation grade exceeding 2. Consistent with these observations, the uniaxial tensile results obtained using the constructed

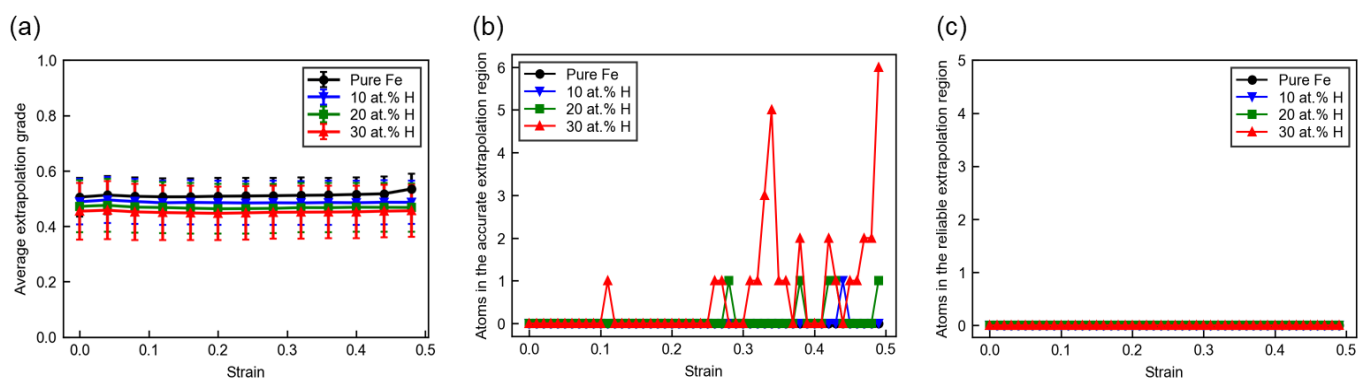

**Fig. 9. Accuracy of the PACE-based machine learning interatomic potential in capturing the deformation of nanocrystals with an average grain size of 4.8 nm and the influence of hydrogen.** Evolution with increasing strain of (a) the mean and standard deviation of extrapolation grade values for all atoms in the nanocrystal, (b) the number of atoms within the "accurate extrapolation" range (grade 1–2), and (c) the number of atoms within the "reliable extrapolation" range (grade ≥2).

MLIP show excellent agreement with those from the MTP, which accurately reproduces the deformation and fracture behavior of nanocrystals containing general GBs. These findings confirm the high reliability and accuracy of the MLIP in simulating uniaxial tension of pure α-Fe nanocrystals.

In the presence of hydrogen, atoms within the "accurate extrapolation" range (extrapolation grades 1–2) were observed, with their number increasing slightly as hydrogen content increased. However, even at the step with the maximum count, only six atoms—representing a mere 0.006% of the entire nanocrystal—fell into this category, an exceedingly small fraction. No atoms exhibited extrapolation grades ≥2, which would necessitate additional labeling in an active-learning procedure. These results indicate that the constructed MLIP delivers highly reliable predictions without requiring further training data.

In the 4.8 nm nanocrystals with hydrogen segregation, significant GB sliding and hydrogen–GB interactions generate a wide variety of local atomic structures, particularly near the boundaries. Plastic relaxation mechanisms, such as dislocation emission, are largely absent, resulting in localized atomic structures at GBs that experience stresses sufficient to initiate cracks. Moreover, the formation and propagation of microcracks at general GBs with hydrogen segregation produce additional relevant atomic configurations. Collectively, these atomic structures closely correspond to those associated with HE involving cracking at general GBs, demonstrating that the constructed

MLIP can reliably analyze even these highly complex systems.

Figure 10 illustrates the uniaxial deformation and fracture behavior of nanocrystals with an average grain size of 19.7 nm, highlighting the effects of hydrogen segregation. Figure 11 presents the evolution with increasing strain of (a) stress, (b) intragranular dislocation density, (c) the number of cracks, (d) crack volume, (e) the number of deformation twins, and (f) the atomic fraction of deformation twins. The stress–strain curves indicate that the maximum stress occurs at a strain of 6–7% (Fig. 11(a)), coinciding with the onset of increasing intragranular dislocation density (Fig. 11(b)). Notably, the peak stress associated with dislocation emission rises with increasing hydrogen content, suggesting that hydrogen segregation suppresses dislocation emission from GBs. This observation is consistent with previous uniaxial tensile analyses of general GB bicrystals performed using BNNP [65].

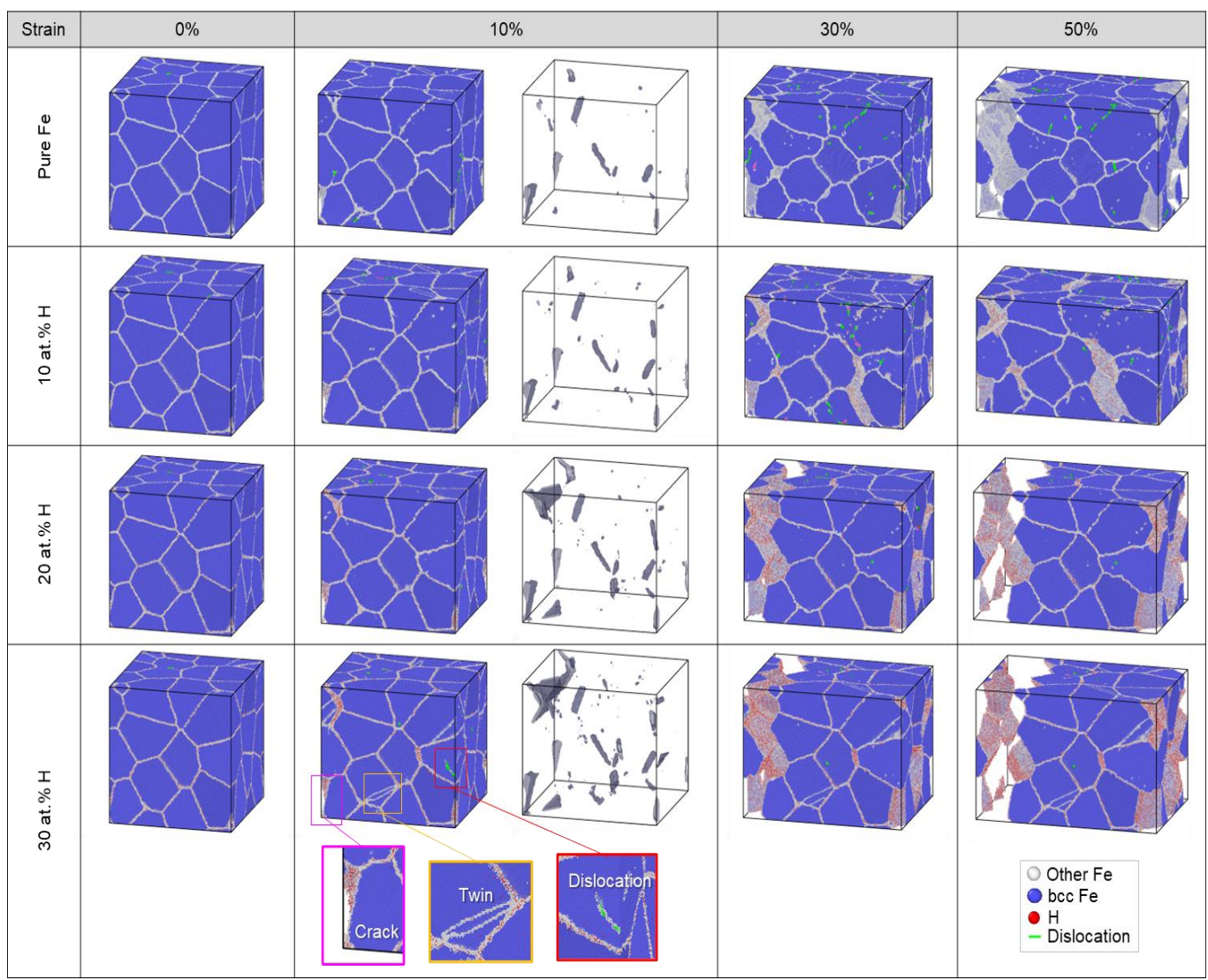

**Fig. 10. Atomic structural evolution of α-Fe nanocrystals with an average grain size of 19.7 nm under uniaxial tension and the influence of hydrogen.** Atomic crystal structures are identified using PTM, and dislocations are visualized via DXA analysis. For 10.0% strain, the crack morphology is shown as determined using a surface mesh.

In the 0–5% strain range, pure α-Fe exhibits a notable deviation from the linear stress–strain relationship, and stress increases with hydrogen segregation (Fig. 11(a)), indicating that GB sliding contributes to deformation. However, the effect of hydrogen is less pronounced than in the 4.8 nm nanocrystals. In the 5–6% strain range, just below the onset of dislocation emission, cracks begin to form, with the strain at crack initiation decreasing as hydrogen content increases (Figs. 11(c),(d)). These observations, consistent with the 4.8 nm nanocrystals, indicate that hydrogen promotes crack nucleation.

Deformation twins emerge near a strain of ~8%, above the peak stress, except in the 20 at.% H case (Figs. 11(e),(f)), with stress levels similar across all hydrogen concentrations (Fig. 11(a)). Compared with pure Fe, both the number and atomic fraction of twins increase at 10 at.% H (Figs. 11(e),(f)), while the intragranular dislocation density is higher in pure Fe (Fig. 11(b)). These results suggest that hydrogen segregation may facilitate deformation twin formation, potentially by indirectly suppressing dislocation emission.

In the high-strain region beyond the peak stress, higher hydrogen content reduces intragranular dislocation density (Fig. 11(b)) while markedly increasing both the number and volume fraction of cracks (Figs. 11(c),(d)). This indicates a synergistic effect in which hydrogen simultaneously suppresses dislocation emission and promotes crack initiation and propagation. Consequently, 19.7 nm nanocrystals with 30.0 at.% H undergo pronounced embrittlement, reaching complete fracture at approximately 20% strain, whereas complete fracture does not occur at 10 at.% H (Figs. 10 and Fig. 11(a)). The presence of numerous microcracks at lower hydrogen concentrations (Fig. 11(c)) likely mitigates stress concentration, impeding the growth of larger cracks required for catastrophic failure (Fig. 11(d)).

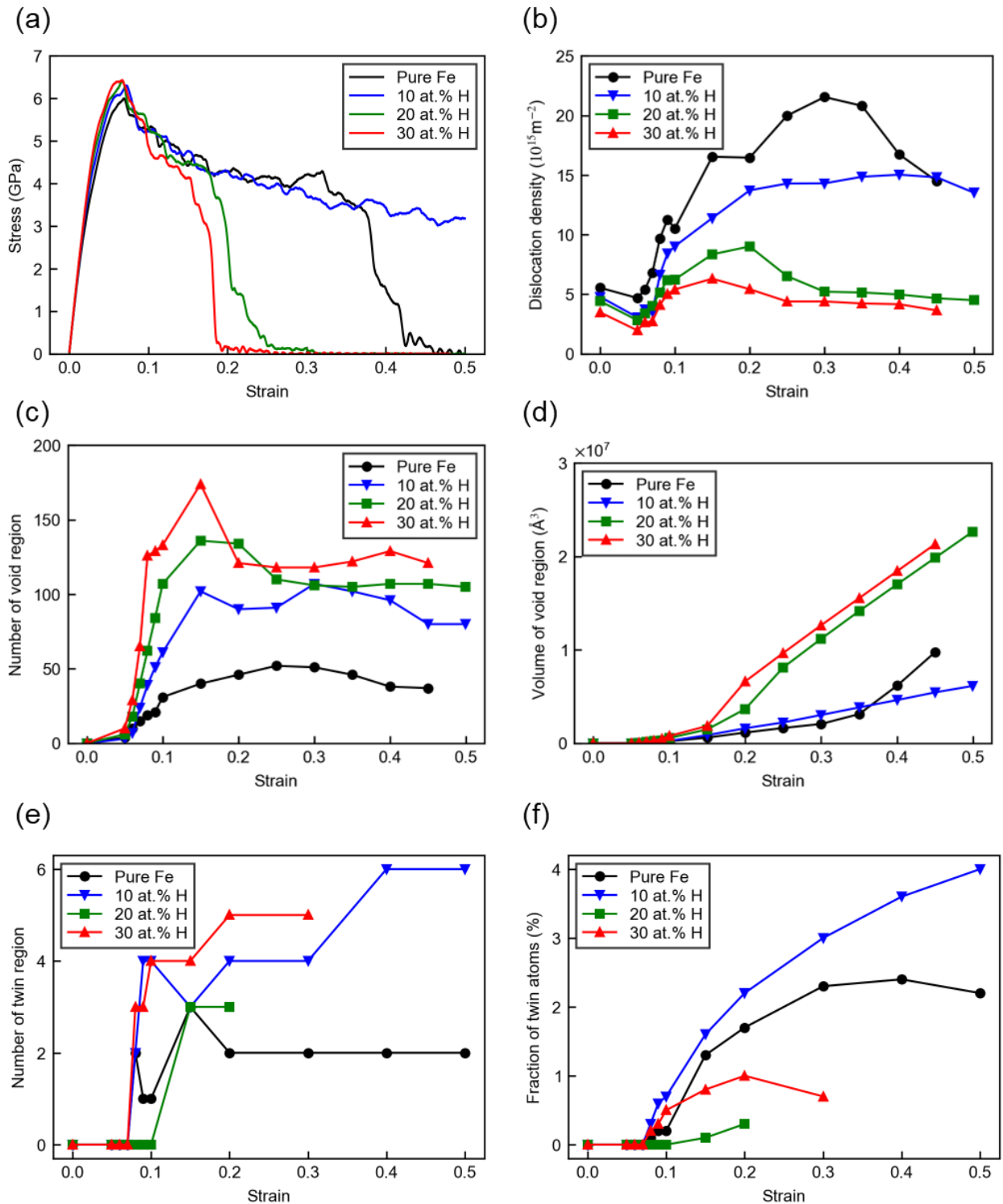

**Fig. 11. Quantitative analysis of atomic structural evolution in α-Fe nanocrystals with an average grain size of 19.7 nm under uniaxial tension and the influence of hydrogen.** Evolution with increasing strain of (a) stress, (b) intragranular dislocation density, (c) number of cracks, (d) crack volume, (e) number of deformation twins, and (f) atomic fraction of deformation twins.

Figure 12(a) presents the evolution of the mean and standard deviation of extrapolation grades for all atoms (approximately 5.4 million Fe atoms and up to 130,000 H atoms) during tensile deformation. Figures 12(b) and (c) show the number of atoms with extrapolation grades in the ranges 1–2 and ≥2, respectively, as functions of strain. The mean extrapolation grade remains approximately constant at ~0.4, similar to the 4.8 nm nanocrystals, corresponding to the "interpolation domain" where high accuracy is expected (Fig. 12(a)). Atoms in the high-accuracy extrapolation domain (grades 1–2) number up to several hundred depending on strain (Fig. 12(b)), with their quantity slightly increasing at low strains for higher hydrogen segregation. Even at the step with the largest number of such atoms, they account for only ~0.005% of the total nanocrystal, an extremely small fraction comparable to that in the 4.8 nm nanocrystals.

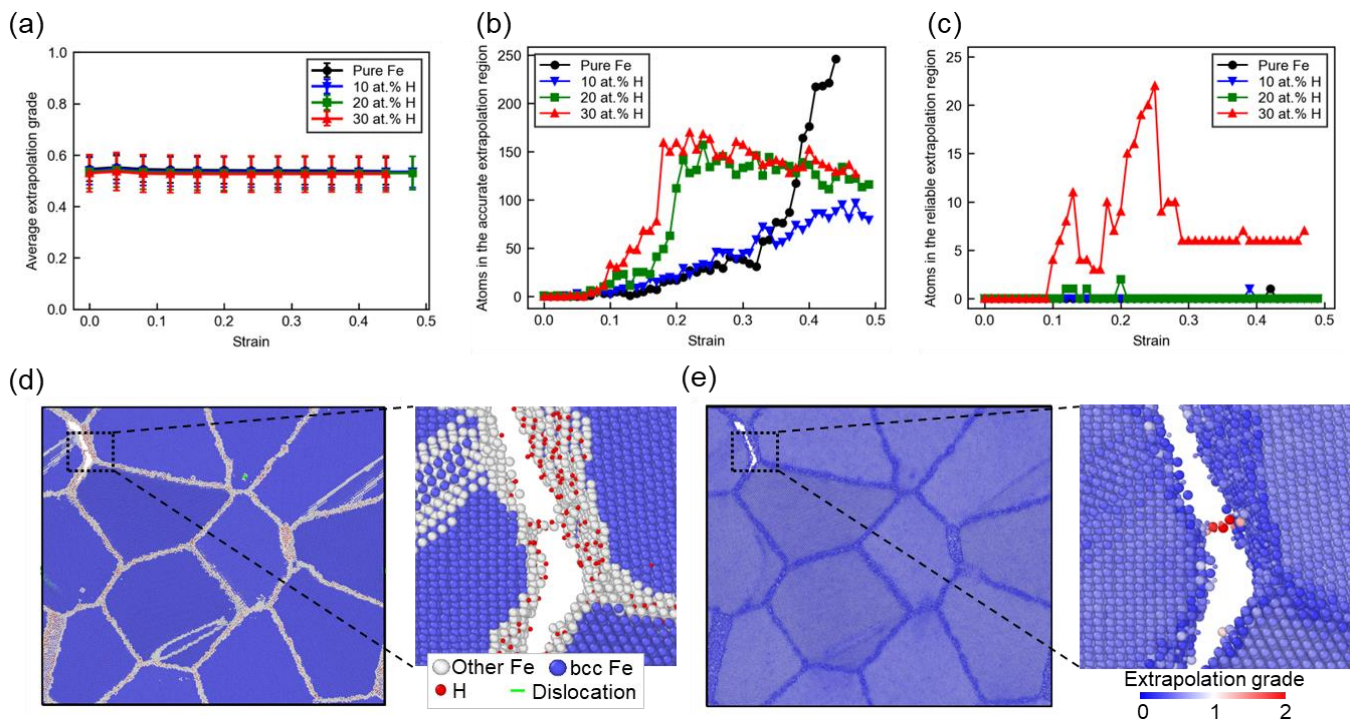

**Fig. 12. Accuracy of the PACE-based machine learning interatomic potential in predicting the deformation of nanocrystals with an average grain size of 19.7 nm and the effects of hydrogen.** Changes with increasing strain in (a) the mean and standard deviation of extrapolation grades for all atoms, (b) the number of atoms within the "accurate extrapolation" range (grades 1–2), and (c) the number of atoms within the "reliable extrapolation" range (grades ≥2). (d) Atomic structure and (e) extrapolation grades for a nanocrystal with 30 at.% H at 10% strain. In (d) and (e), atoms neighboring those with extrapolation grades exceeding 2, which would require additional training data, are also highlighted.

For pure Fe, 10 at.% H, and 20 at.% H, only a few atoms exhibit extrapolation grades ≥2—requiring additional labeling in active learning—throughout the deformation process (Fig. 12(c)). In the 30 at.% H case, up to 22 atoms with grades above 2 appear at the strain step with the highest count.

To evaluate the potential impact of these atoms, Fig. 12(d) and (e) show the local atomic structures surrounding them at the earliest strain (~10%) where they occur. These high-extrapolation-grade atoms are located in constricted regions of the GB immediately prior to complete decohesion, forming quasi-one-dimensional structures with roughly equal proportions of Fe and H. Such atomic configurations differ significantly from bulk structures included in the training data and from configurations observed during the separation of GB-forming grains, justifying the high extrapolation grades. However, these structures are highly localized and appear only immediately before complete decohesion. Considering that over 100 cracks form in total, these cases are rare and are unlikely to significantly affect the overall analysis results.

In summary, for the 19.7 nm nanocrystalline material with hydrogen segregation, the deformation processes encompass dislocation emission from hydrogen-segregated GBs, subsequent dislocation motion along slip planes, and dislocation accumulation, collision, and absorption at the segregated GBs. Consequently, a variety of atomic structures capturing dislocation–hydrogen interactions are present. Furthermore, atomic configurations associated with crack initiation at hydrogen-segregated general GBs of diverse character, as well as their subsequent propagation leading to complete GB fracture, are included. These structures are closely related to HEDE and reflect local atomic configurations central to the hydrogen-enhanced-plasticity-mediated decohesion mechanism [29], such as dislocations accumulating near GBs and their interactions with hydrogen. Because these structures encompass the responses of various GBs under tensile deformation, they may also inform investigations into the crystallographic ease of plastic relaxation of individual GBs, as highlighted in recent studies [24]. Overall, these configurations are highly relevant to atomic processes underlying HE with GB cracking, and the constructed MLIP enables reliable analysis of these complex phenomena. Although the high strain rate used in the simulations precludes observation of hydrogen-enhanced dislocation slip associated with the HELP mechanism, the motion of screw dislocations and their interactions with hydrogen are reproduced with DFT-level accuracy, making the MLIP a valuable tool for studying these processes.

## 3.3. Applications of the findings

In this study, we successfully developed a high-accuracy, computationally efficient Fe–H binary MLIP using a training dataset generated via a concurrent-learning strategy applied to atomic structures relevant to Fe HE. The MLIP accurately reproduces the fundamental properties of α-Fe, lattice defects, and their interactions with hydrogen at DFT-level fidelity. Extrapolation-grade analysis confirms that the MLIP reliably captures the uniaxial tensile deformation and fracture behavior of hydrogen-segregated nanocrystals, including atomic environments associated with hydrogen-induced GB cracking. Moreover, the computational efficiency of the MLIP exceeds that of previous models by more than an order of magnitude. Consequently, the constructed MLIP represents a powerful tool for advancing the understanding of HE mechanisms in steel. Beyond Fe–H systems, the approach used to generate the training data and the adoption of the PACE framework—balancing high accuracy with computational efficiency—can be applied to construct MLIPs for other hydrogen-containing binary metals.

Using the constructed MLIP, we further analyzed the impact of hydrogen segregation at GBs on the uniaxial tensile behavior of nanocrystals. These analyses were performed to validate the MLIP's accuracy for atomic configurations relevant to HE cracking at general GBs. The results revealed suppression of GB sliding, inhibition of dislocation emission, and promotion of deformation twinning due to hydrogen. While these findings are consistent with the high strain rates employed in the simulations, their manifestation at lower, experimentally relevant strain rates remains uncertain. Hydrogen is known to both promote and impede dislocation motion [102]. At high strain and elevated hydrogen concentrations, hydrogen can pin dislocations and restrict their motion [103, 104], whereas at lower strain rates and reduced hydrogen concentrations, it may enhance dislocation mobility [104]. Consequently, dislocation emission from GBs could potentially be facilitated at lower strain rates, underscoring the need for further investigation. The MLIP is well-suited for such studies because it accurately captures dislocation–hydrogen interactions.

Currently, the constructed MLIP is limited to the Fe–H binary system. To gain deeper insights into improving HE resistance in high-strength steels, extension to ternary systems incorporating Fe–H and alloying elements is desirable. Carbon, in particular, is critical for achieving high strength and is expected to significantly influence HE behavior through competitive segregation with hydrogen at GBs [105] and by increasing GB cohesion [106]. Extending the training dataset methodology used here to ternary MLIP construction may lead to a combinatorial explosion of required data; however, this challenge could be addressed using recently developed, advanced concurrent-learning strategies for MLIP construction [107, 108]. Overall, the findings of this study—including the approach for generating HE-relevant training datasets, the development of a high-accuracy, computationally efficient MLIP, and the public release of the resulting potential—are expected to make a significant contribution to the design of high-strength alloys in support of carbon-neutral technologies.

## 4. Conclusions

In this study, we constructed a high-accuracy and computationally efficient Fe–H binary MLIP using a comprehensive training dataset generated via concurrent learning and adopting the PACE format, which offers an optimal balance between fitting accuracy and computational efficiency. The resulting MLIP accurately reproduces the fundamental properties of α-Fe, the formation energies of lattice defects, and the interactions between lattice defects and hydrogen at DFT-level accuracy. Remarkably, it also captures interactions of general GBs, screw dislocations, and edge dislocations with hydrogen, even though these interactions were not explicitly included in the training dataset.

The accuracy of the MLIP was further validated for complex, large-scale systems relevant to HE with GB cracking, including hydrogen-free and hydrogen-segregated nanocrystals containing up to approximately 5 million atoms. These systems encompass diverse atomic structures associated with HE, such as dislocation–hydrogen interactions, crack nucleation at hydrogen-segregated GBs of various character, and progression to complete fracture. Despite this complexity, the MLIP maintained excellent accuracy. Notably, the computational cost is only several tens of times that of EAM, making it more than ten times faster than conventional MLIPs.

Analysis at a high strain rate of $5 \times 10^8$ /s revealed that hydrogen suppresses GB sliding, inhibits dislocation emission from GBs, and promotes both crack initiation and propagation. This study thus presents a robust methodology for constructing high-accuracy, high-speed MLIPs and demonstrates that the resulting potential can significantly advance the understanding and mitigation of HE in various metallic materials.

## CRediT authorship contribution statement

**Kazuma Ito:** Writing – original draft, Methodology, Investigation, Conceptualization, Validation, Writing – review & editing, Supervision, Resources, Project administration, Funding acquisition.

## Declaration of competing interest

There are no known competing financial interests or personal relationships that could have appeared to influence the work reported in this paper.

## Acknowledgments

This work used computational resources of the Supercomputer Fugaku provided by Riken through the HPCI System Research Project (Project ID: hp230272, hp240280, hp250310). This work was partly supported by Accompanying User Support Program (【23Z-03, 23Z-05, 24H1-01】, Support content:【porting of application program, execution performance tuning】) performed by Research Organization for Information Science and Technology.

## Data availability

The machine learning interatomic potential is available online:

https://github.com/KazumaIto0810/Fe-H_PACE .